\documentclass{article}
\usepackage{color}
\usepackage{graphicx}
\usepackage{amsmath, amsthm}
\usepackage{mathtools}
\usepackage{url}
\usepackage{soulutf8}
\RequirePackage[colorlinks,citecolor=blue,urlcolor=blue]{hyperref}
\usepackage{xcolor}
\usepackage{chngcntr}
\usepackage{subfig}
\usepackage{subcaption} 
\usepackage{booktabs}
\usepackage{tikz}
\usepackage{makecell}
\usepackage{hyperref}
\usepackage{algorithm}
\usepackage[noend]{algpseudocode}
\usepackage{setspace}

\usepackage{xfrac}
\definecolor{forest}{rgb}{0.133,0.545,0.133}
\usepackage{multirow}
\usepackage{amsfonts}

\evensidemargin=\oddsidemargin
\usepackage{etoolbox}

\newif\ifabbreviation
\pretocmd{\thebibliography}{\abbreviationfalse}{}{}
\AtBeginDocument{\abbreviationtrue}

\begin{document}
	\newcommand{\bb}{\boldsymbol{\beta}}

	\title{A Personalized Predictive Model that Jointly Optimizes Discrimination and
Calibration}


	\author{Tatiana Krikella\footnote{Tatiana Krikella is the corresponding author and may be contacted at \url{ptkrikel@uwaterloo.ca}.} \hspace{35pt} Joel A. Dubin \bigskip \\ \textit{Department of Statistics \& Actuarial Science} \\ \textit{University of Waterloo, Waterloo, ON, Canada, N2L 3G1}}

	\date{}

	\maketitle

	\begin{abstract}

Precision medicine is accelerating rapidly in the field of health research. This includes fitting predictive models for individual patients based on patient similarity in an attempt to improve model performance. We propose an algorithm which fits a personalized predictive model (PPM) using an optimal size of a similar subpopulation that jointly optimizes model discrimination and calibration, as it is criticized that calibration is not assessed nearly as often as discrimination despite poorly calibrated models being potentially misleading. We define a mixture loss function that considers model discrimination and calibration, and allows for flexibility in emphasizing one performance measure over another. We empirically show that the relationship between the size of subpopulation and calibration is quadratic, which motivates the development of our jointly optimized model. We also investigate the effect of within-population patient weighting on performance and conclude that the size of subpopulation has a larger effect on the predictive performance of the PPM compared to the choice of weight function.
\end{abstract}

		\bigskip


	\maketitle

	\baselineskip=19.5pt


    \section{Introduction}
\label{s:intro}

Models developed in health prediction, including for mortality, are generally good for the average individual, but are not tailored to individuals with unique characteristics. 
This is the motivation behind personalized predictive models (PPMs), where predictions are generated for an index patient using a model developed on a similar subpopulation. 
This area of research is especially valuable in the context of intensive care unit (ICU) prediction models, where it has been suggested that due to the complexity and ambiguity of illnesses in the ICU, a personalized approach should be adopted to both research and practice \cite{maslove2017path}. 

It has been shown that PPMs can lead to better prediction compared to one-size-fits-all models \cite{celi2012database, lee2015personalized}. Consequently, many researchers have proposed algorithms that define a subpopulation of patients that are similar to an index patient $\mathrm{I}$ and train a model using only this subpopulation, ultimately to make a prediction for patient $I$. The most significant differences in these various algorithms are the similarity criterion that is used and the performance measure applied to evaluate the final models. 

One group of algorithms used when fitting PPMs, as noted in a scoping review \cite{sharafoddini2017patient}, are neighbourhood-based algorithms, in which a group of patients similar to patient $I$ is identified to train the PPM.

Hielscher \textit{et al.}'s algorithm for the classification of epidemiological data on the hepatic steatosis disorder is an example of a neighbourhood-based algorithm for fitting a PPM \cite{hielscher2014using}. They use a weighted $k$-nearest neighbour ($k$NN) classifier with Euclidean distance to measure similarity, and choose $k$ to be the value that results in the best average sensitivity, specificity, and accuracy after a two-fold validation. Another neighbourhood-based algorithm, proposed by Ng \textit{et al}, uses a Locally Supervised Metric Learning (LSML) similarity metric, which is based on the Mahalanobis distance measure \cite{ng2015personalized}. To assess the performance of their resulting PPM, they use the area under the receiver operating characteristic curve (AUROC), a commonly calculated measure of model discrimination. A more recent paper by Wang \textit{et al.} also uses Mahalanobis distance to define similarity between patients in their neighbourhood-based algorithm \cite{wang2021study}. They choose the optimal size of subpopulation based on the AUROC, cross-entropy loss and the F1-score, which is a measure of the model's accuracy. 

Non-distance-based similarity metrics used in neighbourhood-based algorithms include a similarity measure proposed by Campillo \textit{et al.} which uses the logical OR-operator \cite{campillo2013improving}. This measure incorporates weights that need to be learned before they are applied to the features. Then, the similarity measure is used to define a subpopulation consisting of the top $M$ similar patients to an index patient, where $M$ is chosen to maximize the AUROC. Another example is the cosine-similarity measure used in an algorithm proposed by Lee \textit{et al.} (2015), where the measure is used to define a subpopulation consisting of the top $M$ similar patients to the index patient, where, again, $M$ is chosen to maximize the AUROC. In later work done by Lee, he suggests that choosing the optimal $M$ value will become important when taking these methods into practice \cite{lee2017patient}.

As mentioned above, a common quantity in choosing the subpopulation size $M$ is the AUROC \cite{sharafoddini2017patient, allam2020patient}, a measure of model discrimination. \textit{Discrimination} is the ability of the predictive model to distinguish between outcome classes. Another measure of model performance is \textit{calibration}, which is the agreement between the observed outcomes and the predictions. For prediction models in general, it has been criticized that calibration is not assessed nearly as often as discrimination \cite{van2019calibration}. This is an issue because poorly calibrated clinical algorithms can be misleading and potentially harmful when making decisions. Van Calster \textit{et al.} also argue that even when calibration is assessed, it is usually done via external validation after the model has been developed via internal validation. Motivated by this issue, Jiang \textit{et al.} propose a prediction algorithm that jointly optimizes the discrimination and calibration of a support vector machine by manipulating the decomposition of an overall performance measure, the \textit{Brier Score} \cite{jiang2012doubly}. The Brier Score is a summary measure that is optimized when the correct probabilities are used, making it a \textit{proper scoring rule} \cite{gneiting2007strictly}. For models that have similar discrimination, a well-calibrated model has a better Brier Score than a miscalibrated model. However, in a recent paper by Assel \textit{et al.}, they claim that the Brier Score does not appropriately evaluate the clinical utility of prediction models \cite{assel2017brier}. 

Through our literature review, we notice that, in general, calibration is not considered when choosing an optimal size of the subpopulation for PPMs. We use this gap in the literature to motivate our work. 

The remainder of the paper will be structured as follows. In Section \ref{s:methods}, we propose an algorithm inspired by Lee et al. to develop a PPM that jointly optimizes both discrimination and calibration. We do this by finding the optimal size of the subpopulation, $M$, that jointly optimizes these performance measures through the use of a proposed loss function, which is an extension of the Brier Score. We use a two-term decomposition of the Brier Score, along with a mixture term, to allow for the flexibility of emphasizing one performance measure over another when training the PPM to find the optimal $M$. Section \ref{s:simulation} contains results from a set of simulation studies which show the relationships between $M$ and various performance measures in addition to a sensitivity analysis on the mixture term of our defined loss function. In Section \ref{s:data}, our
proposed algorithm is tested on data from the eICU Collaborative Research Database, a large multi-center critical care database from hospitals across the United States, to predict mortality of patients with diseases of the circulatory system \cite{pollard2018eicu}.

\section{Methods}
\label{s:methods} 

\subsection{Notation} 

Let there be $n_{train}$ patients in the training data, where each patient $k$ is represented by a Euclidean vector, $\boldsymbol{X}_k$, $k$ = 1, ..., $n_{train}$, in the multi-dimension predictor space defined by $p$ predictors. 
We let $x_h$, $h = 1,...,p$, represent the $n_{train} \times 1$ vector consisting of each patient's value of a given predictor. 
Further, let $\boldsymbol{X}_I$, $I = 1, ..., n_{test}$, be the Euclidean vector representation of the index patient, for whom we would like to make a prediction. Note that the index patient is not included in the training dataset.

The focus of the prediction task in this paper is on binary classification. Thus, we consider a binary outcome. 
The actual outcome for a patient $I$ in the testing dataset is represented by $y_I \in \{0,1\}$. 
Let $M$ be a scalar representing the size of the subpopulation in the training data consisting of similar patients to the index patient.  
Finally, let $p_I^{(M)} \in [0,1]$ represent the predicted probability of the outcome for a patient $I$ found using a model fit on a similar subpopulation of size $M$; we would add further specificity to $p_I^{(M)}$ in the case where there are competing models of the same and/or different type(s), such as $p_{I(logis1)}^{(M)}$, $p_{I(logis2)}^{(M)}$ or $p_{I(RF)}^{(M)}$ for two competing logistic regression models and a random forest model, respectively.

\subsection{Patient Similarity Metric} 

When developing a PPM, one must choose a criterion that will be used to measure similarity between individuals. In this paper, we focus on an unsupervised, one-stage process, where the features being considered for identifying the subpopulation will ideally be identified by subject-matter expert(s), or short of that possibility, through a study of the relevant subject-matter literature. We take this unsupervised approach because in a prediction (future) context, the index patient does not have an outcome label. The similarity between patients $I$ and $k$ is calculated prior to determining a subpopulation which will be used as training data for a PPM to predict the outcome of patient $I$. Another way to think about this is that everyone in the training data has a (similarity) score calculated to measure similarity to the index patient, but only those in the resulting subpopulation, based on $M$, will receive a non-zero weight in the PPM. 

We use the cosine similarity metric as it is bounded, interpretable, and able to handle non-continuous data. Further, it is hypothesized that the cosine metric may be more sensitive to details in patient differences when measuring similarities. \cite{sharafoddini2021identifying} The cosine similarity between patients $I$ and $k$ is shown in Equation (\ref{eq: Cosine1}) where $\bullet$ and $|| \cdot ||$ represent the dot product and the Euclidean vector magnitude, respectively. We label this metric $CSM(\boldsymbol{X}_I, \boldsymbol{X}_k)$.

\begin{equation}
\label{eq: Cosine1}
    CSM(\boldsymbol{X}_I, \boldsymbol{X}_k) = \frac{\boldsymbol{X}_I \bullet \boldsymbol{X}_k}{||\boldsymbol{X}_I||||\boldsymbol{X}_k||}
\end{equation}

Equation (\ref{eq: Cosine1}) is a bounded measure of similarity, with $-1 \leq CSM(\boldsymbol{X}_I, \boldsymbol{X}_k) \leq 1$. $CSM(\boldsymbol{X}_I, \boldsymbol{X}_k) = -1$ corresponds to minimum (opposite) similarity, $CSM(\boldsymbol{X}_I, \boldsymbol{X}_k) = 1$ corresponds to maximum similarity, and a value of 0 refers to orthogonal similarity. 

\subsection{Performance Measures}
\label{ss:performance}

To measure the performance of a PPM, one should measure both the discrimination and calibration of the model \cite{steyerberg_2019}. 
Often, overall performance measures such as the Brier Score, shown in Equation (\ref{eq: Brier}), are used, as the score is affected by both model discrimination and calibration. 

\begin{equation}
\label{eq: Brier}
    BrS = \frac{1}{n_{test}} \sum_{I=1}^{n_{test}} (y_I - p_I^{(M)})^2
\end{equation}

In a binary setting, such as the one we consider in this project, the Brier Score can range from 0 for a perfect model to 0.25 for a non-informative model when the outcome proportion is 50\%. However, the range of the Brier Score changes depending on the outcome proportion, thus making the interpretation of the Brier Score a function of that proportion. A solution to this problem is to scale the Brier Score by its maximum possible score for a realized outcome proportion under a non-informative model \cite{steyerberg2010assessing}. However, in this work, we do not consider the scaled Brier Score as the interpretation of the Brier Score is unnecessary for our purposes, as will be demonstrated below. 

Even when predictive models have high model discrimination, the calibration may still be poor due to statistical overfitting \cite{van2019calibration}. One limitation of the Brier Score is that it may select a model that is miscalibrated with good discrimination over a model that is better calibrated with average discrimination \cite{assel2017brier}. Thus, it may be advantageous to evaluate the discrimination and calibration separately to obtain a more comprehensive assessment of the predictive performance of the model.

A common measure of discrimination is the area under the receiver operating characteristic curve (AUROC), which plots the sensitivity (true positive rate) against 1 – specificity (false positive rate) for consecutive cutoffs for the probability of an outcome. A close-to-perfect discriminated model will have an AUROC close to 1, while a model that cannot distinguish between classes will have an AUROC close to 0.5.
 
Van Calster \textit{et al.} \cite{van2016calibration} have defined a hierarchy of four increasingly strict levels of calibration, referred to as mean, weak, moderate, and strong calibration. We will be using mean, weak, and moderate measures of calibration, as strong calibration is the utopic goal but unrealistic in practice. 

\textit{Calibration-in-the-large} (CITL) is a measure of mean calibration and evaluates whether the observed event rate is equal to the average predicted probabilities, shown in Equation (\ref{eq: CITL}). A CITL value of 0 indicates perfect mean calibration.

\begin{equation}
\label{eq: CITL}
    CITL = \frac{1}{n_{test}} \sum_{I=1}^{n_{test}} |y_I - p_I^{(M)}|
\end{equation}

The \textit{calibration slope} is a measure of weak calibration and is simply the slope of the line that results from plotting the predicted probabilities on the $x$-axis and the observed outcomes on the $y$-axis. A value of 1 indicates perfect weak calibration. Values of the calibration slope below 1 imply that the predicted probabilities are too extreme, while values of the calibration slope greater than one imply that the predicted probabilities are too modest.

The \textit{Integrated Calibration Index} (ICI), valued between 0 and 1, is a measure of moderate calibration \cite{austin2019integrated}, and is the weighted difference between observed and predicted probabilities, where the observations are weighted by the empirical density function of the predicted probabilities. This measure is used to compare the relative calibration of different prediction models, where a lower value of the ICI corresponds to better calibration. An estimate of the ICI is shown in Equation \ref{eq: ICI}, where $\hat{p_I}^{(M)}$ are the smoothed predicted probabilities based on the LOESS calibration curve. This estimate of the ICI intrinsically includes the weights derived from the empirical density function of the predicted probabilities.

\begin{equation}
\label{eq: ICI} 
    ICI = \frac{1}{n_{test}}\sum_{I=1}^{n_{test}}\left|\hat{p_I}^{(M)} - p_I^{(M)}\right|
\end{equation}

\subsection{Mixture Loss Function} 

The optimal size of subpopulation, $M$, is the value which results in a PPM that minimizes some specified loss function. The primary goal of this work is to find the $M$ that results in a PPM with good model discrimination and calibration, while allowing the potential to emphasize one measure over another, if of interest. We capitalize on the fact that the Brier Score, an overall performance measure, can be decomposed into its discrimination and calibration subparts, and invoke a mixture loss term, $\alpha \in [0,1]$, which allows flexibility in emphasizing one performance measure over the other as shown in Equation (\ref{eq: LossFunction}). We use the decomposition discussed by Rufibach, and show details of this decomposition in the Supplementary Material. Specifically, the loss function we use is:

\begin{equation}
\label{eq: LossFunction}
    L^{(M)} = \frac{\alpha}{n_{test}}\sum_{I=1}^{n_{test}}(y_I - p_I^{(M)})(1 - 2p_I^{(M)}) + \frac{1-\alpha}{n_{test}}\sum_{I=1}^{n_{test}}p_I^{(M)}(1-p_I^{(M)}).
\end{equation} 

The first and second terms after the equal sign are related to calibration and discrimination, respectively. Specifically, the first term measures the lack of calibration, with an expectation of 0 under perfect calibration. The second term measures the lack of spread of the predictions. For larger values of $\alpha$, greater emphasis is put on calibration, while for smaller values of $\alpha$, greater emphasis is placed on discrimination. When $\alpha = 0.5$, the loss function is proportional to the Brier Score. 

\subsection{Personalized Predictive Modelling Algorithm} 

We propose an algorithm which sets up a validation procedure to tune the subpopulation size, $M$, for a given value of $\alpha$. We include a hold-out validation step in the algorithm where we estimate the performance of the resulting PPM along with quantifying the uncertainty of these estimates. Step-by-step details of the algorithm are given below. 

\begin{enumerate}
    \item Split the entire population into a $(100q)$\% hold-out validation set and $(100[1-q])$\% training-testing (TrTe) set, where $q\in[0,1]$. We often use $q = 0.2$.
    \item Tune $M$ by following the steps below:
    \begin{enumerate}
        \item Randomly split the TrTe set into distinct training and testing sets using $K$-fold cross-validation (CV).
    \begin{enumerate}
        \item For index patient $I$ in the test set, calculate $d(\boldsymbol{X}_I, \boldsymbol{X}_k)$, for $k=1,...,n_{train}$ in the training set, where $d(\boldsymbol{X}_I, \boldsymbol{X}_k)$ is the similarity metric between the index patient and the $k^{th}$ training set patient.
        \item Consider a grid of $M$ values and tune $M$ through a grid search. 
        \begin{enumerate}
            \item Sort all $d(\boldsymbol{X}_I, \boldsymbol{X}_k)$ in descending order, and the patients corresponding to the top $M$ scores form the subpopulation.
            \item Apply a pre-determined weighting function (eg. uniform, (half-)tri-cube) to each patient in the subpopulation. Note that the (half-)tri-cube weights give those more similar to the index patients a greater weight.
            \item Fit the model of interest, such as logistic regression, random forest, boosted trees, etc., on the subpopulation. 
            \item Average the value of the loss function over the $K$-folds.
    \end{enumerate} 
    \item Repeat the $K$-fold CV $v$ times, where $vK$ is at least $200$. Then, the optimal $M$ is that which has the smallest loss function averaged over the $v$ repeated $K$-fold CV runs. 
    \item The optimal $M$ proportion is obtained by dividing the optimal $M$ by the number of patients in the training set, ie. $P_{optimal} = M/n_{train}$.
    \end{enumerate}
\end{enumerate}
    \item Externally validate the model using the hold-out validation sample to evaluate the performance of all model characteristics (including optimal $M$ proportion and weighting function used to weight patients in the training set).
    \begin{enumerate} 
        \item Draw a bootstrap sample from the validation sample (each the same size as the validation sample).
        \begin{enumerate}
            \item Randomly split the bootstrap sample into an 80\% training sample and 20\% testing sample. 
            \item Set the integer $M = ceiling(n_{val.train}P_{optimal})$ where $n_{val.train}$ is the size of the training set.  
            \item For index patient $I$ in the test set, calculate $d(\boldsymbol{X}_I, \boldsymbol{X}_k)$, for $k=1,...,n_{train}$ in the training set, where $d(\boldsymbol{X}_I, \boldsymbol{X}_k)$ is the similarity metric between the index patient and the $k^{th}$ patient. 
            \item Use the $M$ obtained in step (2) to form the subpopulation
            \item Apply the chosen weighting function to each patient in the subpopulation. 
            \item Fit the model of interest on the subpopulation. 
            \item Measure the performance of the model using, for example, the discrimination and calibration measures discussed in Section \ref{ss:performance}.
        \end{enumerate} 
        \item Repeat the above algorithm with $B$ different bootstrap samples.
        \item In order to quantify uncertainty around the performance measures, fit bias-corrected and accelerated (BCa) bootstrap confidence intervals for each of the performance measures using the $B$ bootstrap estimates.
    \end{enumerate}
    \item Repeat the entire process $Z$ times, ie. repeat the process with $Z$ different validation holdout samples (and, thus, TrTe samples) to view the consistency of the tuning of $M$ (step (1) above) as well as the model performance measures in different hold-out validation samples. We suggest $Z$ to be approximately 10.
\end{enumerate} 

The proposed algorithm above is done for a given $\alpha$, perhaps one that is determined ahead of time, such as $\alpha = 0.5$, where both discrimination and calibration are equally balanced in Equation (\ref{eq: LossFunction}). However, there may be some situations where we want to investigate the performance and resulting $M$ for different scenarios, say a bit more emphasis being placed on calibration over discrimination. We could then implement the above algorithm for a few different $\alpha$ values, perhaps to see how the performance will do as a function of both $\alpha$ and $M$. We will see some of this work in Sections \ref{s:simulation} and \ref{s:data} below.

\section{Simulation Studies}
\label{s:simulation}

We conducted three simulation studies using R \cite{Rprogramming}. We first investigated the relationship between the size of the subpopulation, $M$, and the model discrimination measure, AUROC, to confirm the results of Lee \textit{et al.} that as $M$ increases, the AUROC deteriorates. \cite{lee2015personalized} 
We then investigated the relationship between calibration and the size of subpopulation, $M$. 
Finally, we applied our personalized predictive modelling algorithm to randomly generated data under a number of $\alpha$ values to determine how the value of $\alpha$ affects the optimal $M$ proportion found in the training/testing step of the algorithm. 

\subsection{Setup}
\label{ss:setup}

We considered a sample size of $n=16 000$. 
We simulated $20$ features, where $10$ were binary, $10$ were continuous, and the correlation between any two features was $r = 0.2$. 
We simulated an outcome with a non-linear true outcome model, shown in Equation (\ref{eq: outcomeModel}). This model was passed through an inverse logit function to obtain a probability, $\pi$, as shown in Equation (\ref{eq: inverseLogit}). This probability was then used to simulate a Bernoulli outcome, $y_k$, where $Y_k \sim Bernoulli(\pi)$. Note that the simulated outcome of each index patient in the testing dataset, $y_I$, was generated in the same way.
The prevalence of the outcome was moderate, with values between 0.3 and 0.5 for all simulated datasets. 

\begin{equation}
\label{eq: outcomeModel}
    z = -4x_1 + x_6 - 2x_1x_3 + 3\exp(x_4) - 5\exp(x_2x_8) + \epsilon
\end{equation}

\begin{equation}
\label{eq: inverseLogit}
     \pi = 1/(1+\exp(-z)).
\end{equation}

In practice, this outcome model is usually unknown, thus in our simulation study, we do not fit the model described in Equation (\ref{eq: outcomeModel}) to the index patient to obtain a prediction. Instead, we fit a logistic regression model which considers all $20$ simulated features, but no interactions.

We randomly generated five datasets to test the algorithm, where we conducted 20 repeated 10-fold CV. We tested the algorithm under the following values of $\alpha$: $\{0.475, 0.49, 0.5, 0.58, 0.6, 0.62, 0.75, 0.85, 0.9, 0.99\}$. We did not consider values lower than $\alpha = 0.475$ because the resulting optimal size of subpopulation was too small to apply to the validation set. We discuss this further in Section \ref{s:discuss}.  

In the validation step, we simulated $B=1000$ bootstrap samples. We calculated 95\% bias-corrected and accelerated bootstrap confidence intervals for four performance measures: AUROC, calibration-in-the-large, calibration slope, and the Integrated Calibration Index (ICI). 

\subsection{Relationship Between $M$ and the Performance Measures}

We investigated the relationship between $M$ and the four performance measures.

The results obtained are shown in Figure 1. 
There is a negatively proportional relationship between $M$ and the AUROC in Figure 1(a). As the size of subpopulation increases, the AUROC, hence the discrimination, first has a slight improvement then quickly begins to deteriorate. This confirms the findings of Lee \textit{et al.}

    \begin{figure*}[h]
    \centering
        \subfloat[AUROC]{%
            \includegraphics[width=.38\linewidth]{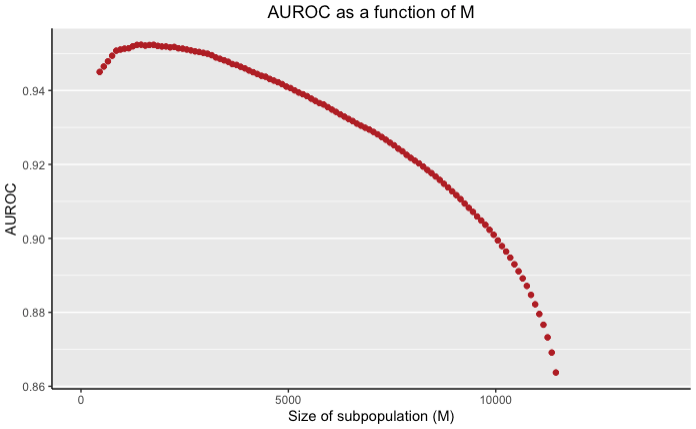}%
} 
        \subfloat[CITL]{%
            \includegraphics[width=.38\linewidth]{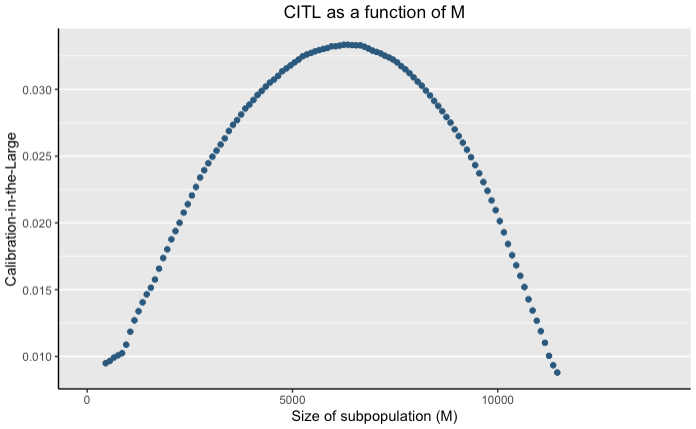}%
}\\
        \subfloat[Calibration Slope]{%
            \includegraphics[width=.38\linewidth]{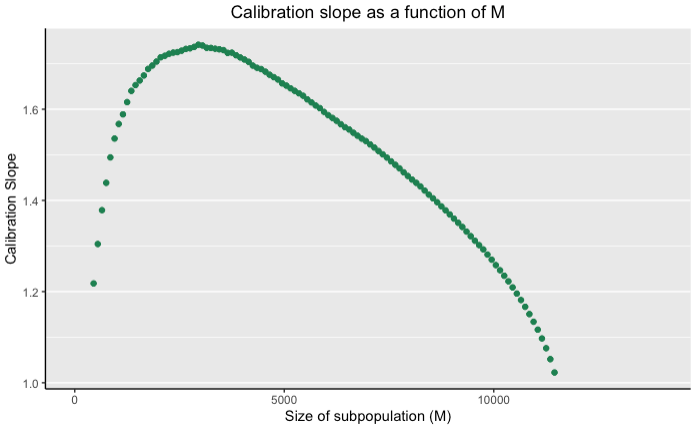}%
}
        \subfloat[ICI]{%
            \includegraphics[width=.38\linewidth]{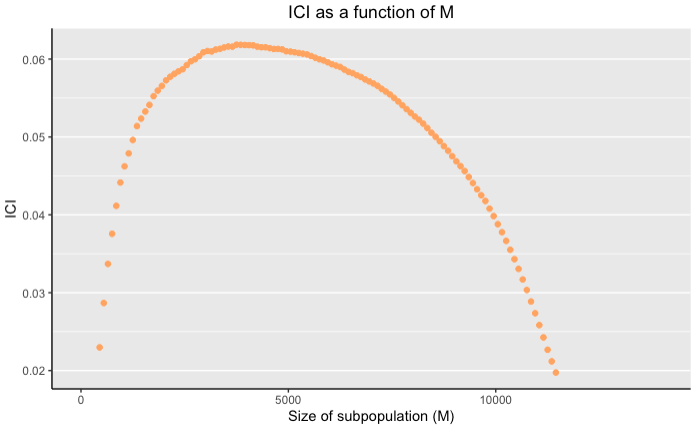}%
}
        \caption{Simulation Study: performance measures as a function of $M$.} 
        \label{fig:simulation_results_wellspecified}
    \end{figure*}
 
The remaining three plots in Figure 1 show the relationship between $M$ and the three calibration measures. In all three plots, the relationship between $M$ and calibration is quadratic: the calibration of the PPM when $M$ is low is good, then as we increase $M$ the calibration deteriorates up to a point, where it then begins to improve. 
Under all three performance measures, the best calibrated PPM is the one which uses all of the available data in the training set. Although, the PPMs using low values of $M$ still are well calibrated, the rate of change of calibration between low $M$ values is much higher than that of higher $M$ values. For instance, Figure 1(d) shows that the difference between the ICI of the PPMs when $M = 450$ and $M = 550$ is $0.0057$, while the difference in the ICI of the PPMs when $M = 11350$ and $M = 11450$ is only $0.0014$. The calibration is more sensitive to changes in $M$ when $M$ is low. 
Under both the calibration slope and the ICI, the calibration of the PPM begins to deteriorate at a much smaller $M$ value compared to the calibration-in-the-large. However, the calibration-in-the-large has a much faster rate of deterioration compared to the other two calibration measures.  

We also tested three different weighting functions to investigate the effect of weighting the similar subpopulation on the performance of the PPMs. We considered uniform weights, (half-)tricube weights \cite{cleveland1979robust}, and a purposely counter-intuitive weighting scheme which gives higher weights to individuals that are the least similar to the index patient. We found that the differences in model performance between the three weighting schemes were negligible, and concluded that the size of subpopulation has a much larger effect on the predictive performance of the PPM compared to the choice of weight function. This is a consistent finding with that in the curve estimation literature, where bandwidth size is much more important than neighbourhood weighting.

\subsection{Applying the Algorithm}
\label{ss:algorithmApplication}

In the last simulation study, we applied the algorithm to randomly generated data and show the results in a variety of plots. 

Figure 2 shows the relationship between the mixture term, $\alpha$, and the resulting optimal $M$ proportion found in the training/testing set of the algorithm. As we increase $\alpha$, the optimal $M$ proportion also increases. The rate of increase is relatively small until $\alpha = 0.75$, where we begin to see a steep slope, especially between $\alpha = 0.75$ and $\alpha = 0.85$. When $\alpha = 0.99$, that is, when there is a very high emphasis on calibration in the loss function, the optimal $M$ proportion is nearly the entire available training population. This algorithm was tested on five different randomly generated datasets and from the plot we see that the results are robust. Many of the datasets resulted in the same optimal $M$ proportions under a similar $\alpha$ value, which is why for some values of $\alpha$ it seems that there are less than five dots. 

\begin{figure}[h!]
    \centering
    \includegraphics[scale=0.38]{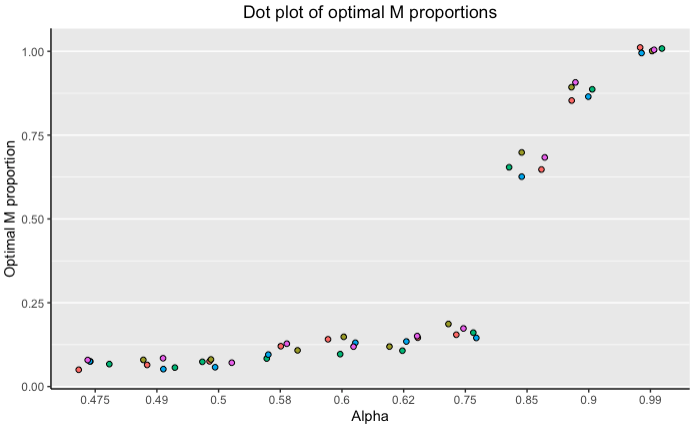} 
    \caption{Dot plot showing the resulting optimal $M$ proportions found in the training step of the algorithm using five different randomly generated datasets.}
    \label{fig:dotplot}
\end{figure}

We then applied the optimal $M$ proportions shown in Figure 2 to the validation step where we assessed the performance of the resulting PPMs. All results are shown via plots in Figure 3. 

\begin{figure*}[h!]
\centering
        \subfloat[AUROC]{%
            \includegraphics[width=.38\linewidth]{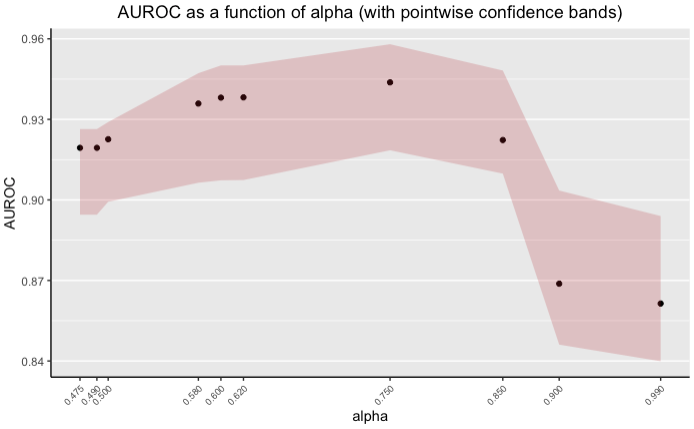}
            }
        \subfloat[CITL]{%
            \includegraphics[width=.38\linewidth]{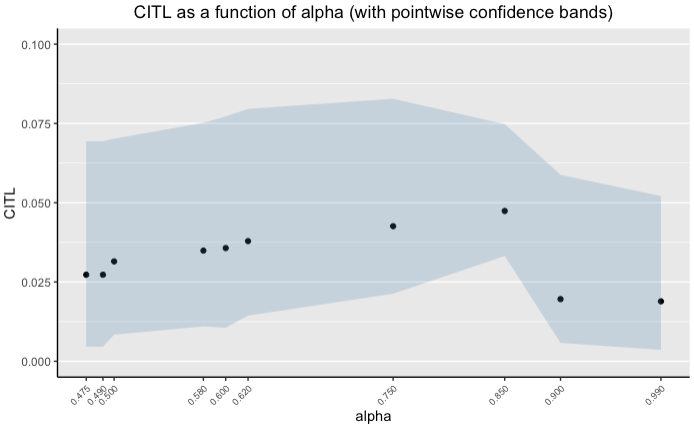}
            }\\
        \subfloat[Calibration Slope]{%
            \includegraphics[width=.38\linewidth]{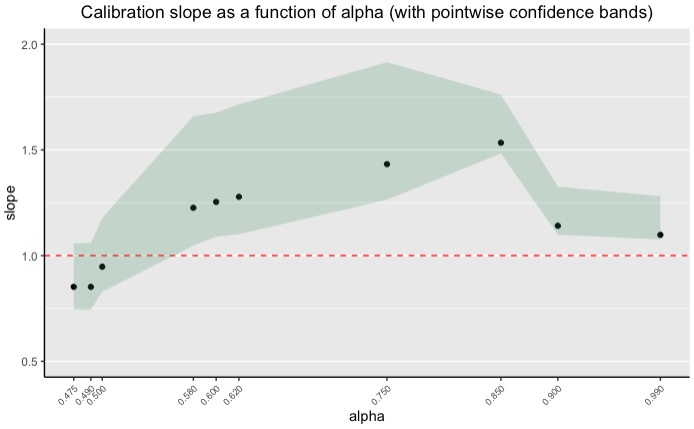}
            }
        \subfloat[ICI]{%
            \includegraphics[width=.38\linewidth]{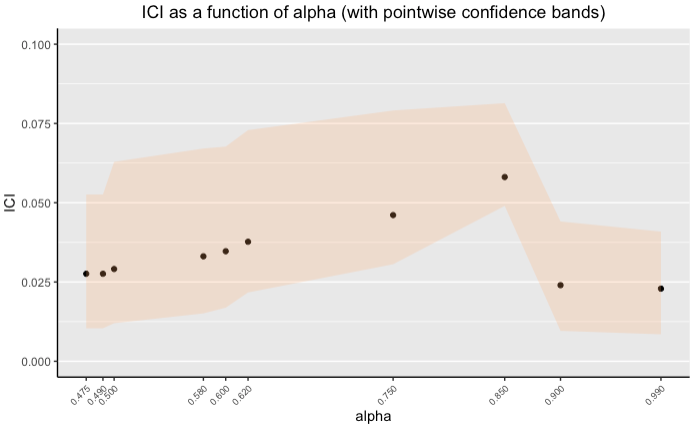}
            }
        \caption{Performance measures as a function of $\alpha$, using the optimal $M$ proportions found in the training/testing stage. The bands shown in each plot represent the 95\% bias-corrected and accelerated (BCa) bootstrap confidence intervals.} 
        \label{fig:validation_results}
    \end{figure*}

Figure 3(a) shows the AUROC as a function of $\alpha$, with the associated pointwise confidence bands. The peak AUROC of 0.944 (0.919, 0.958) was achieved when $\alpha = 0.75$, which corresponds to an optimal $M$ proportion of 0.156. The worst performing PPM is seen under $\alpha = 0.99$, which corresponds to an optimal $M$ proportion of about one. The value of the AUROC under $\alpha = 0.99$ is 0.861 (0.840, 0.894) which is a significant difference. This plot follows the same pattern as shown in Figure 1(a), where we initially see an improvement in discrimination up until a certain point (in this case, $\alpha = 0.75$) where it then begins to deteriorate rapidly. 

The calibration as a function of $\alpha$ is shown in Figures 3(b)-(d). Under the calibration-in-the-large and ICI measures, the best performing PPM is a result of $\alpha = 0.99$, with values of 0.019 (0.004, 0.052) and 0.023 (0.009, 0.041), respectively. Under the calibration slope, the best performing PPM is when $\alpha = 0.5$, which corresponds to an optimal $M$ proportion of 0.078. The value of the calibration slope at $\alpha = 0.5$ is 0.947 (0.829, 1.179). Recall that a calibration slope of one corresponds to perfect weak calibration. Under all three measures of calibration, the worst performing PPM occurs at $\alpha = 0.85$, which corresponds to an optimal $M$ proportion of 0.642. For small and large values of $\alpha$, the PPMs are well calibrated. However, when $\alpha$ is 0.75 and 0.85, these PPMs have poor calibration. This is consistent with the quadratic relationship between $M$ and calibration we showed in Figure 1(b)-(d). Under calibration-in-the-large, the difference between the best and worst calibrated PPMs is 0.029. Under the calibration slope, that difference is 0.586. Finally, under the ICI the difference is 0.035. The PPMs fit in this simulation study are, in general, well-calibrated, so, although appearing small, these differences are particularly meaningful. For instance, a difference of 0.029 in the calibration-in-the-large between two PPMs indicates that the PPM with the worse calibration has an additional 2.9\% of the predicted outcome points that are incorrect compared to the true outcome. In a health setting specifically, when dealing with the mortality of patients, this is not ideal.  

\section{Data Analysis} 
\label{s:data}

We applied our proposed algorithm to a dataset from the eICU Collaborative Research Database \cite{pollard2018eicu} to predict mortality of patients with sepsis. 
The eICU database is a large multi-center critical care database from the United States and is sourced from the eICU Telehealth Program,
which is a model of care where remote providers monitor patients continuously. 
The database consists of 200,859 unit encounters for 139,367 unique patients. 
These patients were admitted between 2014 and 2015 at 208 hospitals located throughout the USA.
The database is deidentified, and includes tables with vital sign measurements, nurse’s notes, severity of illness scoring results, diagnoses, treatment information, care plan documentation, and more. 
One specific table in the eICU database, \textit{vitalPeriodic}, includes data which is interfaced from a bedside monitor at one-minute averages, but is archived in the vitalPeriodic table as five-minute median values. 
Data is publicly available via a PhysioNet repository. 
However, to access, one must complete a training course in research with human subjects.

We only considered patients who were in their first ICU stay of their first hospital visit and had a disease of the circulatory system, identified by their ICD9 code: an official system in the US which assigns codes to various diseases and procedures. We considered only those patients for whom the first hospital stay was identifiable because, although the ordered years of the patient were kept intact, different hospital stays within each year were not ordered chronologically. They must have survived in the ICU for at least 24 hours, but their stay could not exceed 14 days as these patients are outliers, with only 2\% of unique patients having ICU stays lasting more than two weeks in this dataset. An investigation of this group of patients who both survived and stayed in the ICU for over two weeks could be the focus of a future investigation. We removed any patient with missing values in any of the features considered. Further, we removed patients who did not have any vital sign measures in the first 24 hours of their ICU stay. We talk more about handling missing data for this analysis in the Discussion. Figure 4 illustrates the cohort selection from the eICU database.

\begin{figure}[h!]
    \centering
    \includegraphics[scale=0.6]{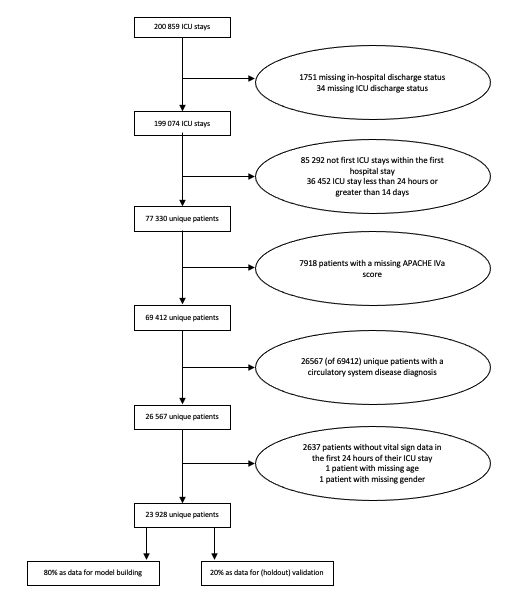} 
    \caption{Flow chart of cohort selection from the eICU database.}
    \label{fig:eICUcohort}
\end{figure}

For each patient, we extracted the following features: 

\begin{itemize}
    \item Admission and demographic data: age, gender (male/female), hospital teaching status, and hospital discharge status (Expired/Alive). 
    \item Vital signs from the first 24 hours after the start time of ICU admission: oxygen saturation (sao2), respiratory rate (respiration), and heart rate.
    \item APACHE IVa score: a severity-of-illness score utilizing demographic information as well information collected within the first 24 hours of a patient's ICU stay.
\end{itemize}

Summary statistics are often used to summarize longitudinal data in analyses of ICU data, such as the vital signs in the case of this data analysis. Such summary statistics might include the mean or median of the repeated measures for a given vital sign, as well as a minimum and maximum value. However, just including summary statistics does not consider the temporal aspect of vital signs. Thus, due to the dense curve-like nature of the vital signs, we use functional principal components (FPCs) to identify each of their dominant modes of variation, specifically for the vital signs in the first 24-hours of a patient's stay in the ICU. We applied FPCA to the vital sign data using the Principal Analysis by Conditional Estimation (PACE) algorithm provided in the R package \textit{fdapace}. \cite{carroll2020fdapace} The PACE algorithm uses eigenfunctions to expand the trajectory of a longitudinal feature for a given patient, as shown in Equation \ref{eq: FPCA}. The first $r$ eigenfunctions ($\hat{\phi}_q(t), q = 1,...,r)$ are used in this expansion, where $\hat{\mu}(t)$ is the estimated mean function of the feature and $\hat{\xi}_{kq}$ represent the FPCs. We extracted the first three FPCs for each of the vital signs, which cumulatively explained at least 95\% of the total variation in the trajectory of each given vital sign; a different dataset for these vital signs, or even different vital signs in this dataset, may have required a different number of FPCs to account for at least 95\% variation.. Our resulting dataset thus consisted of 13 features, including the first three FPC scores for each of the three vital signs.

\begin{equation}
\label{eq: FPCA}
    \hat{X}^r_k(t) =\hat{\mu}(t) + \sum_{q=1}^r \hat{\xi}_{kq}\hat{\phi}_q(t).
\end{equation}

When testing the algorithm, we considered $\alpha = \{0.475, 0.49, 0.5, 0.51, 0.55, 0.6, 0.750, 0.9, 0.95 \}$. We conducted 20 repeated 10-fold CV in the training/testing step of the algorithm. Similar to the simulation study, our outcome in this case is binary, and we fit a logistic regression model to the data. We weighted the subpopulation uniformly. The prevalence of outcome in this dataset is about 0.107, which is much lower than that of our simulation study (about 0.4). Because of this, we calculated the area under the precision recall curve (AUPRC) as a measure of discrimination in addition to the AUROC, as the latter can be misleading when the prevalence of outcome is highly skewed. \cite{davis2006relationship} Specifically, in these cases, the AUROC tends to be too optimistic.

The results from the data analysis are shown in Table 1. The last row of this table displays the model performance when the entire available training data is used to train the PPM, for comparison purposes. The optimal $M$ proportion, in general, increases as $\alpha$ increases. In other words: as we put more emphasis on calibration in our loss function, the size of subpopulation that optimizes discrimination and calibration increases. The one exception to this pattern is $\alpha = 0.475$, which results in a higher optimal $M$ proportion than $\alpha = 0.49$. We found that the differences in the value of the loss function under the first three values of $\alpha$ considered were very close in magnitude, differing by values in the ten thousandths. Thus, for $\alpha = 0.475, \alpha = 0.49$ and $\alpha = 0.5$, any proportion between 0.205 and 0.261 would be acceptable. 

\begin{table}
\caption{Table of results from validating the optimal $M$ proportion found in the training/testing stage of the data analysis. Proportion is optimal proportion found in training step of algorithm. Point estimates are given for each of the performance measures and are in bold. First set of parentheses contain the bootstrap standard error for the estimates. Second set of parentheses contain the BCa bootstrap confidence intervals.}
\centering
\begin{tabular}{|*{7}{c|}}
\toprule
$\alpha$           & Proportion             & AUROC                             & AUPRC         & CITL  & Slope & ICI \\
\midrule
0.475              & 0.261                 & \makecell{\textbf{0.7737} (0.023) \\ (0.7466, 0.7588)}   & \makecell{\textbf{0.5106} (0.054) \\ (0.4459, 0.6248)}               & \makecell{\textbf{0.0211} (0.006) \\ (0.0204, 0.0296)}            & \makecell{\textbf{0.8079} (0.097) \\(0.6456, 0.8985)} & \makecell{\textbf{0.0143} (0.005) \\(0.0051, 0.0222)}    \\  \cline{1-7}

0.49              & 0.205                 & \makecell{\textbf{0.7724} (0.023) \\ (0.7461, 0.7512)}   & \makecell{\textbf{0.5200} (0.054) \\ (0.4704, 0.6274)}               & \makecell{\textbf{0.0212} (0.006) \\ (0.0206, 0.0304)}            & \makecell{\textbf{0.8014} (0.095) \\(0.5987, 0.9243)} & \makecell{\textbf{0.0200} (0.005) \\(0.0111, 0.0327)}    \\  \cline{1-7}

0.5                & 0.205                 & \makecell{\textbf{0.7724} (0.023) \\ (0.7461, 0.7512)}   & \makecell{\textbf{0.5200} (0.054) \\ (0.4704, 0.6274)}               & \makecell{\textbf{0.0212} (0.006) \\ (0.0206, 0.0304)}            & \makecell{\textbf{0.8014} (0.095) \\(0.5987, 0.9243)} & \makecell{\textbf{0.0200} (0.005) \\(0.0111, 0.0327)}    \\  \cline{1-7}

0.51                & 0.534                 & \makecell{\textbf{0.7830} (0.023) \\ (0.7567, 0.7796)}   & \makecell{\textbf{0.5175} (0.054) \\ (0.4551, 0.6258)}               & \makecell{\textbf{0.0219} (0.006) \\ (0.0217, 0.0296)}            & \makecell{\textbf{0.8763} (0.101) \\(0.6435, 0.9911)} & \makecell{\textbf{0.0135} (0.005) \\(0.0051, 0.0203)}    \\  \cline{1-7}

0.55                & 0.871                 & \makecell{\textbf{0.7873} (0.022) \\ (0.7637, 0.7864)}   & \makecell{\textbf{0.5030} (0.053) \\ (0.4311, 0.6194)}               & \makecell{\textbf{0.023} (0.006) \\ (0.0219, 0.0304)}            & \makecell{\textbf{0.8816} (0.103) \\(0.6919, 0.9802)} & \makecell{\textbf{0.0126} (0.005) \\(0.0033, 0.0184)}    \\  \cline{1-7}

0.6              & 0.871                 & \makecell{\textbf{0.7873} (0.022) \\ (0.7637, 0.7864)}   & \makecell{\textbf{0.5030} (0.053) \\ (0.4311, 0.6194)}               & \makecell{\textbf{0.023} (0.006) \\ (0.0219, 0.0304)}            & \makecell{\textbf{0.8816} (0.103) \\(0.6919, 0.9802)} & \makecell{\textbf{0.0126} (0.005) \\(0.0033, 0.0184)}    \\  \cline{1-7}

0.750              & 0.871                 & \makecell{\textbf{0.7873} (0.022) \\ (0.7637, 0.7864)}   & \makecell{\textbf{0.5030} (0.053) \\ (0.4311, 0.6194)}               & \makecell{\textbf{0.023} (0.006) \\ (0.0219, 0.0304)}            & \makecell{\textbf{0.8816} (0.103) \\(0.6919, 0.9802)} & \makecell{\textbf{0.0126} (0.005) \\(0.0033, 0.0184)}    \\  \cline{1-7}

0.9               & 0.871                 & \makecell{\textbf{0.7873} (0.022) \\ (0.7637, 0.7864)}   & \makecell{\textbf{0.5030} (0.053) \\ (0.4311, 0.6194)}               & \makecell{\textbf{0.023} (0.006) \\ (0.0219, 0.0304)}            & \makecell{\textbf{0.8816} (0.103) \\(0.6919, 0.9802)} & \makecell{\textbf{0.0126} (0.005) \\(0.0033, 0.0184)}    \\  \cline{1-7}

0.95               & 0.871                 & \makecell{\textbf{0.7873} (0.022) \\ (0.7637, 0.7864)}   & \makecell{\textbf{0.5030} (0.053) \\ (0.4311, 0.6194)}               & \makecell{\textbf{0.023} (0.006) \\ (0.0219, 0.0304)}            & \makecell{\textbf{0.8816} (0.103) \\(0.6919, 0.9802)} & \makecell{\textbf{0.0126} (0.005) \\(0.0033, 0.0184)}    \\  \cline{1-7}
                    & 1.000                 & \makecell{\textbf{0.7892} (0.022) \\ (0.7625, 0.7857)}   & \makecell{\textbf{0.5085} (0.053) \\ (0.4392, 0.6260)}               & \makecell{\textbf{0.0233} (0.006) \\ (0.0255, 0.0307)}            & \makecell{\textbf{0.9065} (0.103) \\(0.6961, 1.016)} & \makecell{\textbf{0.0120} (0.005) \\(0.0039, 0.0164)}    \\ 
\bottomrule
\end{tabular}
\label{table}
\end{table}

We found that with this dataset, the optimal $M$ proportion was very sensitive to changes in $\alpha$. The optimal $M$ proportion increases by 0.329 when we increase $\alpha$ from 0.5 to 0.51. Then, it increases by 0.337 when we increase $\alpha$ from 0.51 to 0.55. Further, unlike the results from the simulation study, from $\alpha = 0.55$ onwards, the optimal $M$ proportion stays constant at 0.871. 

As we saw in the simulation studies, we notice that in general, as we increase $\alpha$, the discrimination deteriorates and the calibration improves. We do not see this decreasing pattern under the AUROC measure, however we do note that this measure is not reliable when the prevalence of outcome is skewed. Under all $\alpha$ values, the AUROC estimate is greater than the upper bound of its confidence interval, which further supports that the AUPRC is a better measure in this case. We also do not see an increasing pattern under the calibration-in-the-large, but it is the weakest measure of calibration which also sometimes makes it unreliable. An example of its unreliability is shown in the performance of the model that uses the full training data: the point estimate of calibration-in-the-large is not included in the confidence interval. This is why we include measures of weak and moderate calibration.  

The best performing model is seen when $\alpha = 0.51$. Compared to the model with the best discrimination ($\alpha = 0.5$) we sacrifice some discrimination (-0.0025 AUPRC), but we see a large increase in the value of the calibration slope (+0.07), and we also see that this model has better moderate calibration. Further, we see that although the model which uses the full training data has the best calibration, we lose some predictive accuracy through discrimination. This supports what we found in the simulation study. Again, we show that even though researchers can use the Brier Score as the loss function, it may not result in the best performing model.

\section{Discussion}
\label{s:discuss}

While there have been algorithms proposed to fit personalized predictive models (PPMs), there is lack of emphasis in the literature on models that are well-calibrated. We thus proposed an algorithm which fits a PPM using the optimal size of subpopulation that considers both discrimination and calibration. This is done through a novel loss function that we propose which allows for flexibility in emphasizing one performance measure over another, as desired.

We investigated the relationship between the size of subpopulation and calibration and found that the two have a quadratic relationship, with the best calibrated PPMs using the entire population of patients in the training data. However, if one can live with the slightly worse calibration for a smaller $M$ while also wanting better discrimination, then the smaller $M$ will be a better choice that still results in well-calibrated PPMs; alternatively, if calibration is considered much more important than discrimination for a given problem, then the larger $M$ may be the better choice here. Also, the rate of change in calibration between the low values of $M$ is greater than that of higher values of $M$.

The relationship between $M$ and calibration is different from what has been found in the literature regarding discrimination, where, as the size of subpopulation increases, the discrimination deteriorates. Thus, PPMs fit using a low $M$ value will have better model discrimination compared to a PPM fit with a higher $M$ value. We confirmed these results in our simulation study. There were much larger differences in the values of the AUROC between the worst and best performing PPMs compared to what we saw under calibration for both the simulation studies and the data analysis. 

The trade-off between discrimination and calibration when choosing an $M$ value was the motivation behind our proposed algorithm which fits a PPM using an $M$ that considers both performance measures. Our proposed loss function, an extension of the Brier Score, uses the mixture term $\alpha$ to emphasize either performance measure. When the mixture term, $\alpha$, is set to 0.5, then the proposed mixture loss function is proportional to Brier Score. Thus, they attain their minimum at the same value of $M$. Through simulation studies, we found that as we increased $\alpha$, the optimal $M$ proportion found in the training/testing step of the algorithm also increased. In other words, as we put more emphasis on the calibration term in our proposed mixture loss function, the optimal proportion of subpopulation to fit the PPM also increased. However, this rate of increase was not consistent. The largest rate of change we see is between $\alpha = 0.75$ and $\alpha = 0.85$. Before this, the changes in the optimal $M$ proportion between subsequent $\alpha$ values in very gradual. These results were robust across the five randomly generated datasets. In our study, the best performing model was when $\alpha = 0.58$, which means slightly more emphasis was put on calibration compared to discrimination in the mixture loss function. The main purpose of this study was to show that researchers can use the Brier Score as the loss function, but the resulting model will not necessarily be optimal in a general setting. What is optimal will depend on the user and what they care about, whether that is a model that has better discrimination over calibration or vice versa (or a balance between the two), and our next step is to suggest criterion in choosing this $\alpha$ value so the user can can obtain the model performance desired. The $\alpha$ value can also be treated as a tuning parameter in our proposed algorithm, however, this would increase the already intensive computational burden. We were limited in the number of $\alpha$ values below 0.5 we could test, as these $\alpha$ values resulted in very small optimal $M$ proportions. When we applied the optimal $M$ proportion to the validation step, the subpopulation used for training the PPM was not a sufficient size.

To our knowledge, weighting the subpopulation using non-uniform weights had not been investigated, so we tested our algorithm under three different weighting schemes. We found that the choice of $M$ was much more influential on the performance of the resulting model compared to the weighting function used. As stated earlier, this is consistent with smoothing literature, where the choice of bandwidth is more influential than the choice of neighbourhood weighting function.

In our data analysis, we use a dataset from the eICU database to predict mortality of patients with diseases of the circulatory system. Instead of using summary statistics to encapsulate the longitudinal vital sign data, we implemented fPCA so that the temporal aspect of these data was not lost. We applied our proposed algorithm to the data and obtained results consistent with what was found in the simulation study. The largest rate of change in our findings occurred between $\alpha= 0.51$ and $\alpha = 0.55$, however this change is comparable to what we see between $\alpha = 0.5$ and $\alpha = 0.51$. Something interesting to note is that from $\alpha = 0.55$ onwards, the optimal $M$ proportion does not change from 0.871. One reason why the optimal $M$ proportion is not the entire population of patients in the training set for the largest values of $\alpha$ may be due to too much noise in the data, which is common with EHR data. However, this will need to be studied further. The best performing model found in the data analysis corresponded to $\alpha = 0.51$, which used 53.4\% of the full training data as the similar subpopulation. This further supports that using the Brier Score as the loss function may not result in the best performing model. It is important to note that no imputation was performed for variables, such as APACHE IVa, when preparing the data for this analysis. This would be something to focus on in further analyses of these data.

Our study has a few limitations. First, we acknowledge that implementing this procedure for a new index patient will require that the index patient comes from (roughly) the same population of patients on which the $M$ was originally tuned. This may need to be judged on a case-by-case basis. Certainly, we would not want to use an $M$ derived from a different country’s set of hospitals at a different point in time, thus some of the same considerations for externally validating existing prediction models would need to be considered here. Further, our algorithm is very computationally extensive. We employed the \textit{Rcpp} package \cite{Rcpp}, along with parallelization, which sped up the running time significantly, but there is still much room for improvement. If the algorithm can be adjusted to become computationally efficient, other statistical models such as random forest or gradient boosting machines can be tested to see if our results translate to other models, as changing the model that is fit could have an effect on the chosen $M$. Specifically, we wish to investigate how much of an effect the complexity and flexibility of the models fit has on the chosen $M$ value. In addition, in the simulation study and the data analysis we conduct a complete case analysis for simplicity to demonstrate the use of our proposed algorithm, but in practice it is recommended to apply imputation methods to handle the missing data. Implementing such a method, such as multiple imputation, will add another layer of complexity to the algorithm, and how to minimize this added complexity is an area of future work. We also note that the missingness could be be informative regarding similarity between patients, and this idea should also be explored. For example, patients with consistent missing data patterns across different covariates of interest, such as lab tests, could be deemed similar. Another limitation to the studies conducted in this paper is that we considered a relatively small number of features. When working with EHR data, there may be hundreds of features available when fitting prediction models. Thus, a natural next step is to include a variable selection step to our proposed algorithm. Finally, we do not know if the relationships between $M$ and calibration, as well as $M$ and discrimination, that we have shown in this study are specific to the data we considered. Eventually, we want to provide researchers with criterion on choosing an $\alpha$ value, but more work needs to be put into investigating various data scenarios before we can provide more informed advice. 

\clearpage

\bibliographystyle{apalike}

\clearpage 

\appendix
\section{Two-term Brier Score Decomposition}
\label{app1}

\begin{equation}
\begin{array}{cll}%
BrS & = \frac{1}{N}\sum_{k=1}^N [(y_k - p_k)^2] & \\
    & = \frac{1}{N} \sum_{k=1}^N [y_k^2 -2y_kp_k + p_k^2] & \\
    & = \frac{1}{N} \sum_{k=1}^N [y_k -2y_kp_k -p_k +2p_k^2 + p_k -p_k^2] & since \quad y_k \in \{0,1\} \\
    & = \frac{1}{N} \sum_{k=1}^N [y_k(1-2p_k) -p_k(1-2p_k) + p_k(1-p_k)] & \\
    & = \frac{1}{N} \sum_{k=1}^N [(y_k - p_k)(1 - 2p_k) + p_k(1-p_k)] & \\ 
    & = \frac{1}{N} \sum_{k=1}^N [(y_k - p_k)(1-2p_k)] + \frac{1}{N} \sum_{k=1}^N [p_k(1-p_k)] & 
\end{array}
\label{eq:BrierScoreDecomp}
\end{equation}

\end{document}